\documentclass[12pt,onecolumn,a4paper]{article}
\usepackage[super]{cite}
\usepackage{graphicx}
\usepackage{color}
\usepackage{multirow} 
\begin{document}
%%%%%%%%%%%%%%%%%%%%%%%%%%%%%%%%%%%%%%%%%%%%%%%%%%%%%%%%%%%%%%%%%%%
\title{A regular interior solution of Einstein field equations.}

\author{Gabino Estevez-Delgado\\
Facultad de Qu\'imico Farmacobiolog\'ia de la UMSNH\\
Tzintzuntzan No. 173, Col. Matamoros,\\ Morelia Michoac\'an, C.P. 58240, M\'exico.
gabino.estevez@umich.mx, \\
Joaquin Estevez-Delgado\\
Facultad de Ciencias F\'isico Matem\'aticas de la UMSNH \\
 Edificio B, Ciudad Universitaria, \\ Morelia Michoac\'an,  C.P. 58040, M\'exico. joaquin@fismat.umich.mx,\\ 
Modesto Pineda Duran \\
Instituto Tecnológico Superior de Tacámbaro,  \\ 
Av. Tecnológico No 201  Zona el Gigante, C. P. 61650,\\
Tacambaro Michoac\'an, M\'exico.  mpinedad@hotmail.com\\
and  \\
Arthur Cleary-Balderas\\
Facultad de Ingenier\'ia El\'ectrica de la UMSNH \\
Edificio $\Omega$, Ciudad Universitaria, C.P. 58030, \\
Morelia Michoac\'an, M\'exico. arthur.cleary@umich.mx
}

\maketitle
% \pub{Received (Day Month Year)}{Revised (Day Month Year)}
\begin{abstract} 
Starting from the solution of the Einstein field equations in a static and spherically symmetric spacetime which contains an isotropic fluid, we construct a model to represent the interior of compact objects with compactness rate  $u=\frac{GM}{c^2R}<0.23577$. The solution is obtained by imposing the isotropy condition for the radial and tangential pressures, this generates an ordinary differential equation of second order for the temporal $g_{tt}$ and radial $g_{rr}$ metric potentials, which can be solved for a specific function of $g_{tt}$. The graphic analysis of the solution shows that it is physically acceptable, that is to say, the density, pressure and speed of sound are positive, regular and monotonically decreasing functions, also, the solution is stable due to meeting the criteria of the adiabatic index. When taking the data of mass $M=1.44^{+0.15}_{-0.14}M_\odot$  and radius $R=13.02^{+1.24}_{-1.06}km$ which corresponds to the estimations of the star PSR J0030+045 we obtain values of central density $\rho_c=7.5125\times 10^{17} kg/m^3$ for the maximum compactness $u=0.19628$ and of $\rho_c=  2.8411 \times 10^{17} kg/m^3$ for the minimum compactness $u=0.13460$, which are consistent with those expected for this type of stars.
\end{abstract}
\section{Introduction}
\label{sec1}
Describing the interior of the stars and determining their average composition requires many different complementary approaches as are: chemistry, thermodynamics, nuclear physics, particle physics and gravitational physics. And in the case that there are no instabilities generated, once all the nuclear fuel has been used, the star shrinks and, depending of the mass and the stability present, it may form a white dwarf, a neutron star or a quark star \cite{Shapiro,Kippenhahn}.  Of course the description of each one of these stages and their stability is a far more delicate matter which involves a detailed analysis that considers the type of predominant particles in the interior of the star, whether they are electrons, neutrons, or quarks, even when the stars are hybrids. According to the focus of this job, the matter in general is supposed to be described in a satisfactory manner by a perfect fluid and it will not be necessary to give a specific shape of a state equation. We know that, depending on the value of the mass and radius of a star, it may be a white dwarf, a neutron star or a quark star, this will also determine the orders of magnitude of the density, for example densities in the order of $10^{18}kg/m^2$ are typical for neutron stars. As such in this situation, given the high density, it results adequate to describe the interior of the stars by means of Einstein's general relativity theory.  The interior solutions have been approached for over a century, the first of these was constructed for a static and spherically symmetric spacetime with matter from a perfect fluid and incompressible density, known as the interior Schwarzschild solution. Although to start with, the density being constant is an unrealistic requirement, its consequences revealed some differences with the treatment of stellar models in the context of Newton's theory of gravitation. One of these is the compactness relation $u=GM/c^2R<4/9$, where $M$ is the mass and $R$ is the radius, this indicates that it is not possible to have stars with arbitrary mass and radius.
Afterwards it was shown that this relation is not exclusive of this idealized model and that it is present for stars with a monotonically decreasing density function for which the exterior geometry is given by the Schwarzschild solution \cite{Wald,Buch}. And although more that 130 interior solutions with perfect fluid have been published, only a few met the characteristics that makes them physically acceptable. From an analysis done in 1998 on a total of 127 reported solutions, only 16 of these had their density and pressure functions be positive, regular and monotonically decreasing functions and also had a speed of sound that didn't violate the causality condition. And only 9 out of these 16 had a speed of sound which decreases monotonically with the radius \cite{Delgaty,Tolman,Patwardhan,Nariai,Mehra,Kuchowicz,Heintzmann,Adler,Goldman,Matese,Durgapal,Finch}. Although this last requirement is debatable since, for example, for the realistic MIT Bag state equation  $P(\rho)=\frac{1}{3}(c^2\rho-4B_g)$ associated to quark stars, the speed of sound is $v_s^2=\frac{c^2}{3}$, which is not a monotonically decreasing function. The construction of stellar solutions with perfect fluid is an active field although the difficulty in obtaining physically acceptable solutions limits a great number of these. A point which has been explored in relation to these is employing isotropic coordinates \cite{Pant2010,Pant2012a,Pant2014,Molla2022, Murad2014} which has favored the integration of the equations system and its application in the description of stars like HerX1, 4U1538-52, LMC X - 4, SAX J1808.4-3658 \cite{Gangopadhyay}. A particular class of solutions constructed in Schwarzschild's coordinates assumes a metric potential $g_{tt}\!=\!-(1+ar^2)^n$ to this group belong the Tolman IV \cite{Tolman} and Durgapal \cite{Durgapal} solutions, extensions for other values of $n$ a positive integer or negative fractional value \cite{Pant2011a,Maurya2011a,Maurya2011b} have been done, showing that for $n\geq 4$ the solutions that are generated are physically acceptable. Other recent works have addressed the possibility of generating exact solutions with metric potential $g_{tt}\!=\!-\frac{1+ar^2}{1+b^2}$ showing that this functional form is adequate in obtaining physically acceptable solutions and it's consistent with the stars Her X-1 \cite{Estevez2019a,Estevez2019c}, PSRJ0348+0432 \cite{Estevez2019b},  PSR B0943+10 \cite{Estevez2019d},  PSR J0737 -3039A \cite{Estevez2020} and PSRJ1614 2230 \cite{Estevez2021}. 
Motivated by these few last investigation works, in this report we present a new solution to Einstein's equations with perfect fluid with a metric function $g_{tt}$ and its application to the star PSR J0030+045. The structure of this article will be as follows: in the section \ref{seccion2} we present Einstein field  equations for a static and spherically symmetrical spacetime with a perfect fluid and assuming the form of the metric function $g_{tt}\!=\!-
{ {A^2 \left( 1+a{r}^{2} \right)^2 }/\! \left[  { { 1\!+\!
 ( \frac{3}{\sqrt {2} } -1  ) a{r}^{2}}  } \right] }
$  we obtain the solution from the isotropy equation. In section \ref{seccion3} we determine the hydrostatic functions and impose the coupling conditions between the interior and the exterior solutions to determine the integration constants. In section \ref{seccion4} graphic analysis of the solution is done, taking the observational values of the mass and radius of the star  PSRJ0030+045 we determine the physical values of the pressure, density and speed of sound, starting from the graphic analysis and from the data, we show that the solution is physically acceptable. The conclusions and discussion of future works are presented in the section  \ref{seccion5}. 

\section{The field equations and the solution}
\label{seccion2}
The interior geometry of a static and spherically symmetric spacetime can be described through a line element \cite{Weinberg,Schutz} :
\begin{eqnarray}
ds^2=-y(r)^2dt^2+\frac{dr^2}{B(r)}  +r^2 (d\theta ^2+\sin ^2{\theta}d\phi ^2),
\label{elementodelinea}
\end{eqnarray}
where $y,B$ are functions of the radial coordinate $r\leq R$. Einstein equations $G_{\mu\nu}=R_{\mu\nu}-\frac{1}{2}Rg_{\mu\nu}=kT_{\mu\nu}$, (with $R_{\mu\nu}$, $R$ and $g_{\mu\nu}$ the components of the Ricci tensor, the Ricci scalar and the metric tensor respectively), have as source the distribution of matter from a perfect fluid described by the energy-momentum tensor:
\begin{equation}
 T_{\mu\nu}= c^2\rho u_{\mu}u_{\nu}+
 {\it P}(u_{\mu}u_{\nu}+ g_{\mu\nu}), \label{fluido} 
\end{equation}
with $u^\mu$ the four velocity of the fluid,  $\rho$ the energy density  and $P$ the pressure. The non zero components of the Einstein equations are: 
\begin{eqnarray}
kc^2\rho &=&-\frac {B'}{r}+\frac{1-B}{r^2},  \label{rho} \\
k{\it P} &=&\frac{2By'}{ry}-\frac{1-B}{{r}^{2}}, \label{Pr} \\ 
k{\it P} &=&\frac{( ry''+y')B }{ry}+\frac{( ry'+y)B'}{2ry },\qquad \label{Pt}
 \end{eqnarray}
with the derivative in relation to the radial coordinate $r$ denoted by $'$.  Meanwhile the equation of conservation for the energy-momentum tensor $\nabla_\mu T^\mu\,_\nu=0$ implies the Tolman-Oppenheimer-Volkov (TOV) equation \cite{Weinberg,Schutz}: 
\begin{equation}
{\it P}'=-{\frac { \left( {\it P} +c^2\rho  \right) y'}{y }}.\label{conservacion}
\end{equation}
Although this last one is not an independent equation, since it can be obtained from the system of equations (\ref{rho}) - (\ref{Pt}). Being this the set of equations for which we will obtain the solution starting from a function $y(r)$.

\subsection{The solution}
For the integration of the system we propose a metric function $g_{tt}=-y(r)^2$ with the form of $y(r)$ similar, but slightly different, to one employed previously with which it was possible to integrate the system in an adequate manner and which resulted useful to describe compact objects with a compactness rate $u=GM/c^2R\leq 0.2660858316$ \cite{Estevez2019b} . Specifically we have that 
\begin{equation}
y \left( r \right) ={\frac {A \left( 1+a{r}^{2} \right) }{\sqrt {1+
 \left( \frac{3}{\sqrt {2} } -1\right) a{r}^{2}}}},
 \label{y}
\end{equation}
where $A$ and  $a$ are constants. One useful relation for the integration of the system is obtained by replacing $y(r)$ in the difference of the equations (\ref{Pr}) and (\ref{Pt}), leading to:
$$
% \begin{equation}
B' -{\frac {\sqrt {8}[3+\sqrt {2}+4(\sqrt {8}-1)a{r}^{2}+7{a}^{2}{r}^{4}] B  }{( 1+\sqrt {2}a{r}^{2} )  \left( 2+3\sqrt {2}+7a{r}^{2} \right) r}}+{\frac { ( 1+a{r}^{2})  (2+ 3\sqrt {2}+7a{r}^{2}) 
\sqrt {2}}{ ( 1+\sqrt {2}a{r}^{2} )  (3+ \sqrt {2}+7a{r}^{2} ) r}}=0, 
% \nonumber
%\end{equation}
$$ 
after the integration we obtain:
\begin{eqnarray}
B \left( r \right) &=&1+{\frac { \left( 31 -22\sqrt {2}  \right)  \left( 3\,\sqrt {2}+2+7\,a{r}^{2} \right) ^{3}a{r}^{2}}{343\, \left( 1+a\sqrt {2}{r}^{2} \right) ^{3}} \left[  C+\ln  \left( {\frac {\sqrt {2}+3+7\,a{r}^{2}}{3
\,\sqrt {2}+2+7\,a{r}^{2}}} \right)  \right] }\qquad \nonumber\\
&+&
{\frac { \left( 22-17\sqrt {2} \right)  \left( 732\sqrt {2}+ 1832+7\left( 555\sqrt {2}+251 \right) a{r}^{2}+4606{a}^{2}{r}^{
4} \right) a{r}^{2}}{2303\left( \sqrt {2}+2a{r}^{2} \right) ^{3}}
}
\label{B},
\end{eqnarray}
$C$ is the constant of integration.
\section{Hydrostatic functions and physical conditions}
\label{seccion3}
Once we know the metric functions we will proceed to determine the hydrostatic functions. Replacing the functions $y(r)$ and $B$ given by the equations (\ref{y}) and (\ref{B}) in the equations (\ref{rho}) and (\ref{Pr}) we determine the density and the pressure:
\begin{eqnarray}
k{c}^{2}\rho \left( r \right) &=& {\frac { 3\left( 2\,\sqrt {2}+6+
 \left( -4+15\,\sqrt {2} \right) a{r}^{2}+14\,{a}^{2}{r}^{4} \right) 
 \left( 1-B \left( r \right)  \right) }{ \left( 3\,\sqrt {2}+2+7\,a{r}
^{2} \right)  \left( \sqrt {2}+2\,a{r}^{2} \right) {r}^{2}}}
\nonumber\\
&&-{
\frac { 14\left( 12\,\sqrt {2}-13+7\,a{r}^{2} \right) {a}^{2}{r}^{2}}{
 \left( 3\,\sqrt {2}+2+7\,a{r}^{2} \right)  \left( \sqrt {2}+2\,a{r}^{
2} \right)  \left( \sqrt {2}+3+7\,a{r}^{2} \right) }},
 \\
kP(r) &=& {\frac { 2\left( 6\,\sqrt {2}-3+7\,a{r}^{2}
 \right) a}{ \left( 1+a{r}^{2} \right)  \left( 3\,\sqrt {2}+2+7\,a{r}^
{2} \right) }}
\nonumber\\
&&
-{\frac { \left( 3\,\sqrt {2}+2+3\, \left( 5\,\sqrt {2}+
1 \right) a{r}^{2}+21\,{a}^{2}{r}^{4} \right)  \left( 1-B \left( r
 \right)  \right) }{ \left( 1+a{r}^{2} \right)  \left( 3\,\sqrt {2}+2+
7\,a{r}^{2} \right) {r}^{2}}}.
\label{Pfinal}
\end{eqnarray}
In these equations the expression $(1-B)/r^2$ appears, however, it is regular when $r=0$ as it can be seen from the equation (\ref{B}). Another important relation to determine if the solution is physically acceptable is the speed of sound, since it is required that the speed of sound in the model does not violate the causality condition. In this case by means of the chain rule we obtain the speed of sound: 
$$
\frac{v^2(r)}{c^2}=\frac{1}{c^2}\frac{\partial P(\rho)}{\partial \rho}={\frac {
 \left( S_{{1}} \left( r \right) B \left( r \right) + \left( 3\,\sqrt 
{2}+2+7\,a{r}^{2} \right) ^{2} \left( 1+a{r}^{2} \right) ^{2} \right) 
S_{{2}} \left( r \right) }{ \left( S_{{3}} \left( r \right)  \left( 
\sqrt {2}+3+7\,a{r}^{2} \right) ^{2}B \left( r \right) +S_{{4}}
 \left( r \right)  \right)  \left( 1+a{r}^{2} \right) ^{2}}},
$$
where 
\begin{eqnarray}
S_{{4}} (r)&=&( 1+a{r}^{2} )  \left[ 15\sqrt {2}+150+7( 34+9\sqrt {2}) a{r}^{2}+98{a}^{2}{r}^{4} \right] \left[3\sqrt {2}+2+7a{r}^{2} \right] ^{2}\!,\nonumber \\
S_{{1}} (r)&=& \left( \sqrt {2}-4 \right)  \left( \sqrt {2
}+3+7\,a{r}^{2} \right)  \left( 2+\sqrt {2}+ \left( 5\,\sqrt {2}-1
 \right) a{r}^{2}+3\,{a}^{2}{r}^{4} \right) ,\nonumber \\
S_{{3}} (r)&=&3\, \left( \sqrt {2}-4 \right)  \left( 10\,
\sqrt {2}+30+3\, \left( -4+15\,\sqrt {2} \right) a{r}^{2}+14\,{a}^{2}{
r}^{4} \right) ,\nonumber \\
S_{{2}} (r)&=& \left( 6\,\sqrt {2}-3+7\,a{r}^{2} \right) 
 \left( \sqrt {2}+3+7\,a{r}^{2} \right)  \left( \sqrt {2}+2\,a{r}^{2}
 \right) .\nonumber 
 \end{eqnarray}
\subsection{Criteria for physical acceptability}
Obtaining a solution to Einstein's equations is not a guarantee that said solution is physically acceptable, there are many solutions that are not physically acceptable \cite{Delgaty} due to the fact that they do not comply with certain properties. In the following we will mention the requirements that must be satisfied, some of these will be applied directly and others will be shown in a graphical manner in the following section \cite{Estevez2019b}. 

%\noindent
One of the conditions that must be met, is the {\bf regularity of the geometry} when approaching the center. Which can be expressed in an algebraic manner, through the Kretschmann scalar, given its extension, it is enough with showing that the metric coefficients around $r=0$ are of the form $\alpha+\beta r^2+O(r^4)$. The expansion of $B(r)$ and $y(r)$  in the proximity of $r=0$ gives us:
$$
y \left( r \right) =A \left( (1-{\frac { \left( 3\,\sqrt {2}-6
 \right) a}{4}}{r}^{2}-{\frac { \left( 30\,\sqrt {2}-41 \right) {a}^{2
}}{16}}{r}^{4}+O \left( {r}^{6} \right) ) \right)
, 
$$
$$
B \left( r \right) =1+{\frac {2\,a \left(  \left( 17\,\sqrt {2}-26
 \right)  \left( C+\ln  \left( {\frac {\sqrt {2}+3}{3\,\sqrt {2}+2}}
 \right)  \right) +41\,\sqrt {2}-80 \right) {r}^{2}}{49}}+O \left( {r}
^{4} \right),
$$
besides the regularity, the geometry must be {\bf absent of any event horizon}, this property is easier to demonstrate through a graphic analysis and it will be analysed in the following section. \\
\noindent
{\bf The density and pressure must be finite, positive and monotonically decreasing} as functions of the radial coordinate. That is to say, for $r\in(0,R)$, $\rho'<0$ and $P'<0$ (condition that will be analysed graphically)  and in the center they must have their maximum value. which implies the following set of inequalities:
\begin{equation}
k{c}^{2}\rho ( 0 ) ={\frac {6\,a \left(  \left( 132\,\sqrt 
{2}+193 \right)  \left( 2\,C-\ln  \left( 2 \right)  \right) +3006\,
\sqrt {2}+4286 \right) }{6713\,\sqrt {2}+9506}}>0,
\label{rho0}
\end{equation}
\begin{equation}
kP( 0 ) =1/49\,a \left(  \left( 17\,\sqrt {2}-26
 \right)  \left( 2\,C-\ln  \left( 2 \right)  \right) -65\,\sqrt {2}+
134 \right) 
>0,
\label{P0}
\end{equation}

\begin{equation}
\rho''(0) =-{\frac {5 {a}^{2} \left[ 6\left( 195-103\sqrt {2}
 \right)  \left( 2 C-\ln 2  \right) +3819+6257 \sqrt {2} \right] }{49\, \left( \sqrt {2}+3 \right) ^{3}}}<0,
\end{equation}
\begin{equation}
\!\!P''(0)=-{\frac {3 {a}^{2} \left[ 2\left( 113-72\,\sqrt {2}
 \right)  \left( 2\,C-\ln2  \right) +2759-1248\,
\sqrt {2} \right] }{49\, \left( \sqrt {2}+3 \right) ^{2}}}
 <0.
 \label{P20}
\end{equation}
In addition to these inequalities the solution satisfies $\rho'(0)=$ and $P'(0)=0$, that together with the inequalities (\ref{rho0})-(\ref{P20}) implies that $r=0$ is a maximum for the functions $\rho$ and $P$. 
The {\bf causality condition in the center} of the star requires that it satisfies
\begin{equation}
\!\!0\leq\frac{v(0)^2}{c^2}% \frac{1}{5}
={
\frac {3[24\,\sqrt {2}+4\,C-2\,\ln \, 2 +\,55\,]}{5[96\sqrt {2}+12C-6\ln 2   +121]}}\leq 1.\label{v0}
\end{equation}
Combining the previous equations we can determine inequalities for the constants  $(C,a)$, in particular
forming $k \left( \rho \left( 0 \right) {c}^{2}+3\,\Pr \left( 0 \right)  \right) =9\, \left( 2-\sqrt {2} \right) a
>0$, from where we obtain that $a>0$. \\
\noindent
The constants $C$ and $W$ which appear in the metric functions are determined by imposing that {\bf the interior and exterior geometry on the surface of the star $r=R$ are joined in a continual manner and that the pressure is zero on the surface}. The exterior geometry is described by the exterior Schwarzschild solution:
\begin{eqnarray}
ds^2\! &\!=\!&-\left(1-\frac{2GM}{c^2r}\right)\,dt^2+\left(1-\frac{2GM}{c^2r}\right)^{-1}dr^2
+\,r^2 (d\theta ^2+\sin ^2{\theta}\,d\phi ^2),
\quad r\geq R,
\nonumber
 \end{eqnarray}
where $M$ represents the total mass inside the fluid sphere.  When we impose $P(R)=0$, from the equation (\ref{Pfinal}) we obtain $C$:  
\begin{eqnarray}
C&=&-\ln  \left( {\frac {\sqrt {2}+3+7\,w}{3\,\sqrt {2}+2+7\,w}}
 \right) +{\frac { \left( 25\,\sqrt {2}+47 \right)  W_1  }{ W_2  \left( 3\,\sqrt {2}+2+7\,w \right) ^{2}}},
\label{Ccondicion}
\end{eqnarray}
where $w=aR^2$, $
W_2=%\left( 
822
\,\sqrt {2}+548+822\, \left( 5\,\sqrt {2}+1 \right) w+5754\,{w}^{2}
$ and 
$$
W_1=3286
\sqrt {2}+4048+( 8149\sqrt {2}+18623) w+7( 2757\sqrt {2}+1075) {w}^{2}+13426{w}^{3}.
$$
Meanwhile from the continuity of the component $g_{tt}$ in $r=R$ it results:
\begin{equation}
{A}^{2}={\frac { \left( 3\,\sqrt {2}-2 \right)  \left( 3\,\sqrt 
{2}+2+7\,a{R}^{2} \right) ^{2}}{ 14 \left[ 3\,\sqrt {2}+2+3\, \left( 5\,
\sqrt {2}+1 \right) a{R}^{2}+21\,{a}^{2}{R}^{4} \right]  \left( 1+a{R}
^{2} \right) }}.
 \label{yR}
\end{equation}
The continuity of $g_{rr}$ in $r=R$ determines the value of the compactness as function of $w$:
\begin{equation}
u(w)=\frac{GM}{c^2R}=\frac{1}{2}(1-B(R))= {\frac { \left( 6\,\sqrt {2}-3+7\,w \right) w}{3\,\sqrt {2}+2+3\,
 \left( 5\,\sqrt {2}+1 \right) w+21\,{w}^{2}}}.
 \label{u}
\end{equation}

\noindent
The rest of the conditions require of a graphical analysis and these correspond to the {\bf  Energy conditions}:

- The Strong Energy Condition: $ c^2\rho+3P\geq 0$, $ c^2\rho+P\geq 0$ or

- The Dominant Energy Condition: $\rho\geq 0$  and $ c^2\rho\geq |P|$\\
And the  {\bf  Stability condition}, a configuration of static and spherically symmetrical matter is stable if it satisfies the relativistic condition for the adiabatic index:
$$
\Gamma=\frac{P+c^2\rho}{c^2P}\frac{dP}{d\rho}> \frac{4}{3} \qquad \forall \;r\in[0,R]
$$

\section{Graphic representation of the solution}
\label{seccion4}
From the graphic analysis of the functions of density, pressure, speed of sound and adiabatic index we obtain that the function which restricts the values of the parameter $w\leq w_0=0.90378$ is that of the adiabatic index, specifically for values of $w > w_0$ the adiabatic index $\gamma (0) <4/3$, which implies that the solution will be unstable. 
This maximum value $w_0$ through the equation (\ref{u}) allows us to obtain the maximum permissible compactness value for the compactness $u\leq u_0= 0.23577$. Although the solution is physically acceptable for the compactness values  $u\leq u_0$ in the graphic analysis we will focus on the particular case of the star PSR J0030+0451 with estimates of mass $M=1.44^{+0.15}_{-0.14}M_\odot$  and radius $R=13.02^{+1.24}_{-1.06}km$, obtained through the study of the X-ray emission by means of the NICER (Neutron star Interior Composition Explorer) telescope from the international space station\, \cite{PSRJ00300451}. The graphic representation will be done in terms of the dimensionless variable $x=r/R$ and the dimensionless functions associated to the physical quantities of density $kc^2R^2 \rho$,  pressure $kR^2 P$ and speed of sound $v^2/c^2$. The values of compactness that were chosen for the graphic analysis are $u_{max}=0.19628$,
$u=0.18086$,  $u=0.16545$,  $u=0.15003$ and   $u_{min}=0.13460$,  where  $u_{max}$ it's associated with the maximum mass  $M=1.59 M_\odot$ and the minimum radius $R=11.96km$ $u=u_{max}=$, meanwhile  $u_{min}$ is obtained by taking the minimum mass $M=1.3 M_\odot$ and the maximum radius $R=14.26km$. 
\begin{figure}[htb!]
\begin{minipage}[t]{0.43\linewidth}
\centering
\includegraphics[width=6.3 cm]{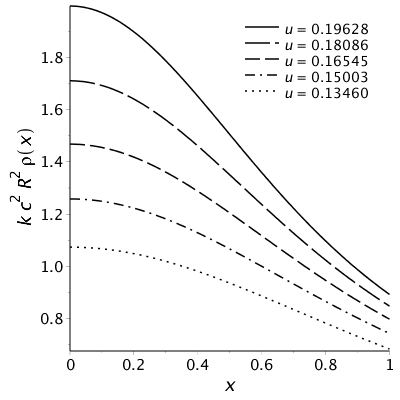}
\end{minipage}
\hspace{0.9cm} 
\begin{minipage}[t]{0.43\linewidth}
\centering
\includegraphics[width=6.3 cm]{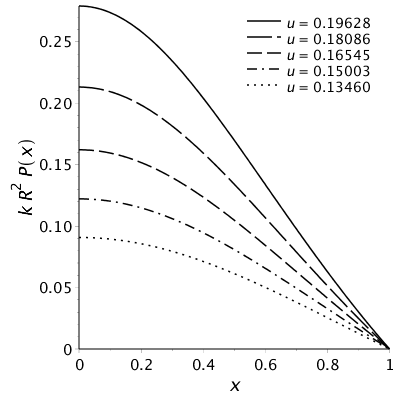}
\end{minipage}
\caption{Graphic representation of the density and the pressure for the different values of compactness from the star PSR J0030+0451.}
\label{dendsidadypresion}
\end{figure}

\begin{figure}[htb!]
% \hspace{0.1cm} 
\begin{minipage}[t]{0.43\linewidth}
\centering
\includegraphics[width=6.3 cm]{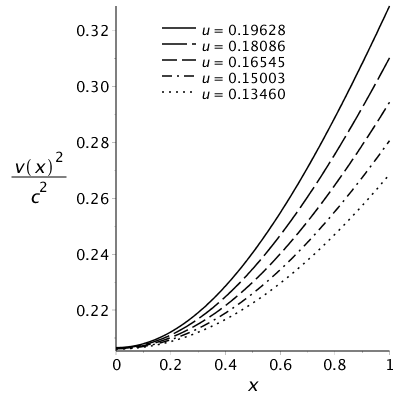}
%	\caption[The density EXO 1785-248.]{Behaviour of the ordinary density for the compactness $u= 0.21691$.} \label{KmQ6crho}
\end{minipage}
\hspace{0.9cm} 
\begin{minipage}[t]{0.43\linewidth}
\centering
\includegraphics[width=6.3 cm]{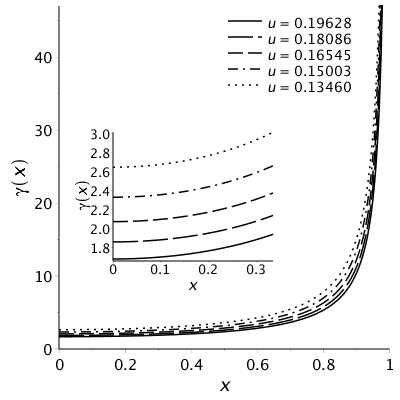}
%	\caption[ \hspace 2cm The density EXO 1785-248.]{Representation of the quintessence density for the compactness $u= 0.21691$.} \label{KmQ6crhoq} 
\end{minipage}
\caption{Graphic representation of the speed of sound and the adiabatic index for the different compactness values of the star PSR J0030+0451.}
\label{velocidadeindice}
\end{figure}

\begin{figure}[htb!]
% \hspace{0.1cm} 
\begin{minipage}[t]{0.43\linewidth}
\centering
\includegraphics[width=6.3 cm]{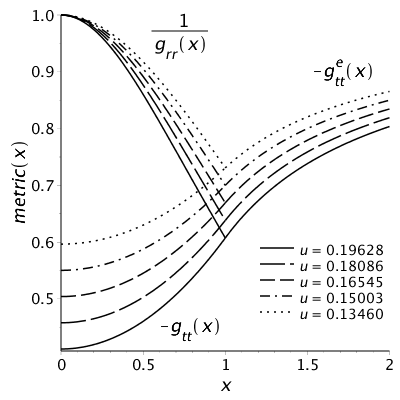}
%	\caption[The density EXO 1785-248.]{Behaviour of the ordinary density for the compactness $u= 0.21691$.} \label{KmQ6crho}
\end{minipage}
\hspace{0.9cm} 
\begin{minipage}[t]{0.43\linewidth}
\centering
\includegraphics[width=6.3 cm]{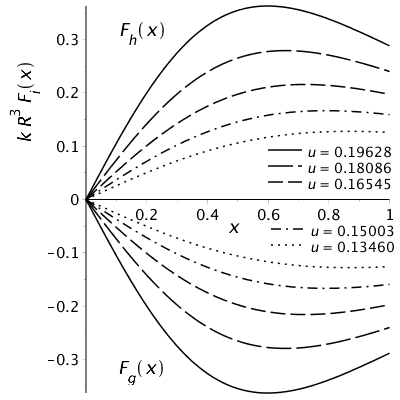}
%	\caption[ \hspace 2cm The density EXO 1785-248.]{Representation of the quintessence density for the compactness $u= 0.21691$.} \label{KmQ6crhoq} 
\end{minipage}
\caption{In the graph of the left side we present the behaviour of the metric coefficients from the interior and exterior geometry, meanwhile in the graph of the right side it´s shown the behaviour of the forces in the interior of the star.}
\label{metricayfuerza}
\end{figure}

\noindent
In the figure \ref{dendsidadypresion} we show the behaviour of the density and pressure for different values of compactness which is obtained from the estimates in base to observations. The graphics show that the density and pressure are positive and monotonically decreasing functions, their values diminish as the compactness value decreases, appearing in a more noticeable manner the difference between the values of the density or the pressure in the center of the star, we also observe that the pressure is zero on the surface. From the figure \ref{dendsidadypresion} we observe that the Strong Energy Condition is satisfied, since both the density and the pressure are positive. Also we have that for a specific value of $u$ the value of the density is much greater than that of the pressure ($c^2\rho>P$), which implies that the Dominant Energy Condition is also satisfied. 
From the figure \ref{velocidadeindice}, graphic on the right, we observe that the causality condition is met, since $0.2 c^2<v^2<0.34 c^2$ and that the speed of sound is lower for lower compactness values, with maximum values on the surface. The stability of the solution is guaranteed by the adiabatic index, the left graph in the figure \ref{velocidadeindice}, with $\gamma$ being a monotonically increasing function, the lowest value of the adiabatic index occurs in the center of the star for the maximum compactness $u_{max}$ as such the set of compactness values that is being analysed satisfies $\gamma>1.6843>4/3$. 
The absence of an event horizon and the continuity of the geometry on the surface of the star is shown in the figure  \ref{metricayfuerza}. In addition,in the graph on the right side of the figure \ref{metricayfuerza}, we graph the forces present (the gravitational force $F_g$ and the hydrostatic force $F_h$), identified by means of the Tolman-Oppenheimer-Volkoff (TOV) equation 
$$
-{\frac { \left( {\it P_r} +c^2\rho  \right) y'}{y }}-{\it P_r}'=0,\qquad 
\Rightarrow \qquad 
F_g(r)=-{\frac { \left( {\it P_r} +c^2\rho  \right) y'}{y }},\quad
F_h(r)=-{\it P_r}'.\qquad
$$
In the figure \ref{metricayfuerza} we can observe the attractive effect of the gravitational force $F_g$ countered by the hydrostatic repulsive force.
%\begin{eqnarray}
%F_g(r)&=&-{\frac { \left( {\it P_r} +c^2\rho  \right) y'}{y }}\qquad
%F_h(r)=-{\it P_r}'\qquad
%F_a(r)=\frac{2}{r}(P_t-P_r).
%\end{eqnarray}

\section{Discussion and conclusions}
\label{seccion5}
The graphic analysis has allowed us to show that the solution presented satisfies every requirement which makes it physically acceptable and although the graphic analysis was done considering estimated values of mass and radius for the star PSR J0030+045, a similar behaviour is present for other values of compactness, as long as $u\leq 0.23577$. To confirm that the behaviour of the solution is not only graphically compatible but also that the orders of magnitude which are obtained from the model are compatible with those expected, on the tables \ref{Tabla1} and \ref{Tabla2} we report the values of density, pressure, speed of sound and adiabatic index in the interior for the case of maximum \ref{Tabla1} and minimum compactness \ref{Tabla2}. 
\begin{table}[h]
\caption{Interior behavior of the physical values of the density, pressure, speed of the sound and adiabatic index for the PSR  J0030+0451, with  $R=11.96 km$ and $M=1.59 M_\odot$, $u_{max} = 0.19628$.}
{ \begin{tabular}{|c|c|c|c|c|}
\hline
$r(km)$&$\!\!\rho\left(10^{17}\frac{kg}{m^3}\right)\!\!\!$&$\! P(10^{33} Pa)$ & $v^2(c^2)$& $\gamma$  \\ \hline 
     0.&  7.5125 &  9.4371 &  0.20654 &  1.6843  \\  \hline
1.1960 &  7.4155 &  9.2563 &  0.20795 &  1.7052   \\  \hline
2.3920 &  7.1398 &  8.7360 &  0.21214 &  1.7702   \\  \hline
3.5880 &  6.7211 &  7.9258 &  0.21905 &  1.8887   \\  \hline
4.7840 &  6.2131 &  6.9019 &  0.22857 &  2.0775   \\  \hline
5.9800 &  5.6621 &  5.7429 &  0.24055 &  2.3719   \\  \hline
7.1760 &  5.1146 &  4.5238 &  0.25474 &  2.8428   \\  \hline
8.3720 &  4.5970 &  3.3040 &  0.27094 &  3.6581   \\  \hline
9.5680 &  4.1283 &  2.1264 &  0.28887 &  5.3282   \\  \hline
10.764 &  3.7155 &  1.0202 &  0.30824 &  10.395   \\  \hline
11.960 &  3.3547 &       0 &  0.32880 & $\infty$  \\  \hline  
\end{tabular} }
\label{Tabla1}
\end{table}
\begin{table}[h]
\caption{Interior behavior of the physical values of the density, pressure, speed of sound and adiabatic index for the  PSR J0030+0451 star, with  $R=14.26 km$ and $M=1.3 M_\odot$, $u_{min}= 0.13460$.}
{ \begin{tabular}{|c|c|c|c|c|}
\hline
$r(km)$&$\!\!\rho\left(10^{17}\frac{kg}{m^3}\right)\!\!\!$&$\! P(10^{33} Pa)$ & $v^2(c^2)$& $\gamma$  \\ \hline 
    0  &  2.8411 &  2.1586 &  0.20605 &  2.6431  \\  \hline
1.4260 &  2.8246 &  2.1275 &  0.20669 &  2.6723  \\  \hline
2.8520 &  2.7760 &  2.0362 &  0.20873 &  2.7656  \\  \hline
4.2780 &  2.6978 &  1.8894 &  0.21203 &  2.9321  \\  \hline
5.7040 &  2.5965 &  1.6944 &  0.21667 &  3.2001  \\  \hline
7.1300 &  2.4774 &  1.4588 &  0.22249 &  3.6176  \\  \hline
8.5560 &  2.3462 &  1.1923 &  0.22956 &  4.2876  \\  \hline
9.9820 &  2.2093 &  0.90520 &  0.23777 &  5.4501  \\  \hline
11.408 &  2.0715 &  0.60586 &  0.24708 &  7.8340  \\  \hline
12.834 &  1.9374 &  0.30200 &  0.25734 &  15.081  \\  \hline
14.260 &  1.8055 &  0       &  0.26859 & $\infty$ \\  \hline
\end{tabular} }
\label{Tabla2}
\end{table}
From the tables  \ref{Tabla1} and \ref{Tabla2} it can be noticed that the orders of magnitude of the density and pressure are also those expected for the star  PSR  J0030+0451. With which we can conclude that the model obtained is physically acceptable and useful to represent stars with compactness $u\leq 0.23577$. Another relevance of the solution constructed is that it can be useful as seed for obtaining new physically acceptable solutions \cite{Boonserm} in which we consider the contribution of the presence of electric charge \cite{carga} or from an anisotropy factor \cite{Ovalle4}, as well as in the determination of new solutions in alternative gravitational theories \cite{alternativas}, investigations that could be developed in future works.

  \newpage
\section*{Acknowledgments} 
We appreciate the facilities provided by the Universidad Michoacana de San Nicolás de Hidalgo and the CIC -UMSNH during the realization of this investigation as well as the CONACYT for the support given.

\end{document}